\documentclass{article}

\usepackage[margin=1in]{geometry}
\usepackage{amsmath,amssymb,amsfonts}
\usepackage{graphicx}
\usepackage{booktabs}
\usepackage{caption}
\usepackage{subcaption}
\usepackage{hyperref}
\usepackage{natbib}
\usepackage{siunitx}
\usepackage{microtype}
\newcommand{\ECVAupper}{\ensuremath{\mathrm{ECVA}^{\mathrm{upper}}_s(\varepsilon)}}
\newcommand{\CCVAupper}{\ensuremath{\mathrm{CCVA}^{\mathrm{upper}}_s(\varepsilon)}}
\newcommand{\WWRDelta}{\ensuremath{\Delta^{\text{WWR}}_s(\varepsilon)}}

\usepackage{float}    
\usepackage{placeins} 
\usepackage{etoolbox} 

\AtBeginDocument{%
  \pretocmd{\section}{\FloatBarrier}{}{}%
  \pretocmd{\subsection}{\FloatBarrier}{}{}%
  \pretocmd{\subsubsection}{\FloatBarrier}{}{}%
}

\hypersetup{
  colorlinks=true,
  linkcolor=blue,
  citecolor=blue,
  urlcolor=blue
}

\title{Environmental CVA with KL-Robust Wrong-Way Risk}
\author{Takayuki Sakuma\footnote{e-mail: tsakuma@soka.ac.jp. }\\
Faculty of Economics and Business Administration, Soka University}
\date{\today}

\begin{document}
\maketitle

\begin{abstract}
Although climate- and nature-related scenario analysis is increasingly important in finance, operational implementations remain limited for translating long-horizon environmental scenarios into counterparty credit-risk measures used in pricing and regulatory capital. We propose an environmental valuation-adjustment framework for CVA with three components: (i) a scenario-to-credit translation that maps environmental scenario drivers into hazard rates; (ii) nature-specific tail generators that quantify model risk in scenario generation; and (iii) a distributionally robust wrong-way-risk bound based on Kullback–Leibler (KL) divergence. We compute climate CVAs using transition scenarios and nature CVAs using biodiversity indicators. Our results show that nature CVAs can vary materially across alternative ecosystem generators, highlighting an additional source of model uncertainty.
\end{abstract}


\section{Introduction}
Environmental financial risk is increasingly addressed through two related supervisory agendas: climate risk and nature-related risk. Climate risk already enters supervisory scenario analysis, and NGFS pathways are widely used as public reference inputs for supervisory stress testing, ICAAP-style scenario analysis, and capital planning; recent CCVA work also adapts long-horizon climate scenarios to valuation-adjustment settings \citep{NGFSPhaseVPortal,NGFSPhaseVUserGuide,kenyon2021ccva}. Nature-related risk is moving along a different but rapidly developing track: disclosure and governance frameworks are advancing through initiatives such as the TNFD, supervisory workstreams are emerging through the NGFS and FSB, and biodiversity risk is increasingly entering the finance literature \citep{tnfd2023recommendations,ngfs2023conceptual,fsb2024nature,giglio2023biodiversityrisk}. Operational methods that translate nature-scenario outputs into counterparty-credit inputs used in pricing and regulatory capital, however, remain limited and heterogeneous \citep{fsb2024nature,unepfi2023nature,oecd2023supervisory}.

This paper addresses this implementation gap by developing an operational framework that maps long-horizon environmental scenarios into counterparty credit valuation adjustments. For climate risk, we build on climate-change valuation adjustment (CCVA) approaches that map long-horizon scenario drivers into hazard-rate multipliers \citep{kenyon2021ccva}. First, we extend the same basic logic to a nature framework driven by public biodiversity scenario outputs. This introduces an additional layer that is less prominent in the climate framework: \emph{scenario-generation uncertainty}.
Even conditional on a fixed policy path, downstream ecosystem outcomes can be nonlinear, generator-dependent, and heavy-tailed; we therefore treat alternative ecosystem simulations as \emph{tail generators} and study how they deform the NCVA distribution.
Second, in both the climate and nature frameworks, we treat wrong-way risk (WWR) as dependence uncertainty.
Rather than imposing a single joint dependence model for exposure and default, we start from an independence benchmark and compute worst-case CVA over a Kullback--Leibler (KL) neighborhood around that benchmark, following \citet{GlassermanXu2014} and \citet{GlassermanYang2018}.
The KL radius $\varepsilon$ is interpreted as an explicit dependence/model-risk budget.

The remainder of the paper is organized as follows.
Section~\ref{sec:method} sets out the common CVA framework and robust WWR construction.
Section~\ref{sec:results} reports the climate benchmark setup and results. Section~\ref{sec:nature} introduces the nature-specific translation and reports the benchmark and case-study results.
Section~\ref{sec:discussion} discusses implications and limitations, and Section~\ref{sec:conclusion} concludes.

\section{Methodology}\label{sec:method}

\subsection{Overview and notation}
Let $s$ index environmental scenarios and let $ref$ denote the reference scenario. Because this section is common to both the climate and nature applications, we use the generic label \emph{environmental CVA} (ECVA) here; Section~\ref{sec:results} specializes it to climate CVA (CCVA), and Section~\ref{sec:nature} specializes it to nature CVA (NCVA).
Under the independence benchmark between market exposures and default risk, we define the scenario-relative ECVA as
\begin{equation}
\mathrm{ECVA}^{\mathrm{ind}}_s \;=\; \mathrm{CVA}^{\mathrm{ind}}_s - \mathrm{CVA}^{\mathrm{ind}}_{ref}
\label{eq:ecva_ind}
\end{equation}
and define the robust scenario-relative ECVA as
\begin{equation}
\mathrm{ECVA}^{\mathrm{upper}}_s(\varepsilon) \;=\; \mathrm{CVA}^{\mathrm{upper}}_s(\varepsilon) - \mathrm{CVA}^{\mathrm{upper}}_{ref}(\varepsilon),
\label{eq:ecva_upper}
\end{equation}
where $\mathrm{CVA}^{\mathrm{upper}}_s(\varepsilon)$ denotes the KL-robust upper bound on scenario-$s$ CVA under a KL radius $\varepsilon$ around the independence benchmark.
Because \ECVAupper\ is defined as the difference between two scenario-specific upper bounds, it is interpreted as a decomposition term rather than as an upper bound on $\mathrm{ECVA}^{\mathrm{ind}}_s$.
Therefore the \WWRDelta\ defined below can be negative if the reference scenario is distorted more than scenario~$s$ under the worst-case joint distribution allowed by the KL budget.

Finally, we define the robust WWR, denoted \WWRDelta, as a difference-in-differences,
\begin{equation}
\Delta^{\text{WWR}}_s(\varepsilon)
=
\mathrm{ECVA}^{\mathrm{upper}}_s(\varepsilon)
-
\mathrm{ECVA}^{\mathrm{ind}}_s
\label{eq:wwr_did}
\end{equation}
and the robust WWR component is measured relative to the same reference path as the ECVA baseline. By construction,
\[
\mathrm{ECVA}^{\mathrm{upper}}_s(\varepsilon)=\mathrm{ECVA}^{\mathrm{ind}}_s+\WWRDelta
\]
and the WWR term is interpreted as an incremental model-risk buffer around the independence benchmark.

\subsection{Independence CVA}
We separate the calculation into three objects: the discounted loss on default, the discrete CVA identity under the independence benchmark, and the scenario-dependent hazard curve.

Let $\tau$ denote the counterparty default time and let $T$ denote the CVA horizon.
For a fixed recovery rate $R$, discounted loss is
\begin{equation}
L = \mathbf{1}_{\{\tau\le T\}}\, DF(0,\tau)\,(1-R)\, EAD(\tau),
\label{eq:loss}
\end{equation}
where $DF(0,t)$ is the discount factor and $EAD(t)$ is exposure at default. When we allow scenario-dependent recovery in the nature CVA, $R$ in Eq.~\eqref{eq:loss} is replaced on the grid by the recovery path $R_s(t_i)$.

Under the baseline independence benchmark, scenario-$s$ CVA is $\mathrm{CVA}^{\mathrm{ind}}_s=\mathbb{E}[L]$.
We discretize default on the same grid as exposures and include an explicit no-default outcome with probability $S_s(T)$.
A standard discrete approximation then yields
\begin{equation}
\mathrm{CVA}^{\text{ind}}_s
=
(1-R)\sum_{i=1}^{n} DF(0,t_i)\,\mathrm{EPE}(t_i)\,\left(S_s(t_{i-1})-S_s(t_i)\right),
\label{eq:cva_discrete}
\end{equation}
where $\{t_i\}_{i=0}^n$ is the simulation grid, expected positive exposure $\mathrm{EPE}(t_i)=\mathbb{E}[\max(V(t_i),0)]$ is computed from the exposure simulation, and $S_s(t)$ is the scenario survival curve implied by the hazard term structure.

We fit a one-factor Hull--White (HW1F) model with deterministic shift to the observed discount curve at $t=0$, simulate the mark-to-market value of an interest-rate swap, and compute positive exposure $E(t)=\max(V(t),0)$ and EPE.

On the credit side, a baseline credit curve is calibrated from a credit-spread measure to a baseline intensity $\lambda_0$.
Scenario $s$ then changes only the hazard term structure through a hazard multiplier.
Let $\mathcal{K}$ denote the chosen driver set (in the climate benchmark, GDP and carbon price).
For each $k\in\mathcal{K}$, let $x_{s,k}(t)$ denote the scenario path of driver $k$ and let $\beta_k$ denote its translation elasticity.
Fixing a base year $t_0$, we define an \emph{absolute} multiplier of log-ratio form,
\begin{equation}
m_s^{\text{abs}}(t)=\exp\left(\sum_{k\in\mathcal{K}} \beta_k \log\frac{x_{s,k}(t)}{x_{s,k}(t_0)}\right).
\end{equation}
We then form a \emph{reference-relative} multiplier $m_s(t)=m_s^{\text{abs}}(t)/m_{ref}^{\text{abs}}(t)$ and finally set the scenario hazard curve to $\lambda_s(t)=m_s(t)\lambda_0(t)$.
Thus the environmental scenarios change CVA through scenario-dependent default probabilities, while the exposure paths remain common across scenarios.

\subsection{KL-robust WWR}
For each scenario $s$, the baseline measure $P$ is the independence benchmark generated by (i) simulating exposure paths under the market model and (ii) sampling $I$ independently from the hazard-implied default distribution.

We then consider alternative joint measures $Q$ on the same space that are absolutely continuous with respect to $P$ and satisfy a KL (relative-entropy) constraint,
$D_{\mathrm{KL}}(Q\|P)=\mathbb{E}_Q[\log(dQ/dP)]\le \varepsilon$.
Following \citet{GlassermanXu2014}, standard relative-entropy duality yields the convex dual representation
\begin{equation}
\sup_{Q: D_{\mathrm{KL}}(Q\|P)\le\varepsilon} \mathbb{E}_Q[L]
=
\inf_{\eta>0}\left\{\eta\varepsilon+\eta \log\mathbb{E}_P\left[\exp\left(\frac{L}{\eta}\right)\right]\right\}.
\label{eq:kl_dual}
\end{equation}

In Monte Carlo terms, Eq.~\eqref{eq:kl_dual} is read as a \emph{reweighting} of the baseline loss sample.
If $L^j$ denotes the $j$th simulated loss under the independence benchmark and $p_j$ its baseline probability (typically $p_j=1/N$ under ordinary Monte Carlo), then for a given dual parameter $\eta$ the weighted probabilities are
\[
q_j(\eta)=\frac{p_j\exp(L^j/\eta)}{\sum_{\ell} p_{\ell}\exp(L^{\ell}/\eta)}.
\]
At the optimizer $\eta^\star$ of the one-dimensional dual problem, the worst-case measure $Q^\star$ is obtained immediately from these normalized exponential weights. Higher-loss realizations are upweighted until the KL budget $\varepsilon$ is exhausted.

We proceed in three steps.
First, we simulate pathwise positive exposures $E_p^+(t_i)=\max(V_p(t_i),0)$ on the grid $\{t_i\}_{i=0}^n$.
Second, for a given hazard curve, we sample default intervals independently from the hazard-implied default distribution and construct discounted loss samples
\[
L_p=(1-R)\,DF(0,t_{I_p+1})\,E_p^+(t_{I_p+1}),
\]
where $I_p$ is the sampled default-interval index.
Third, we solve the one-dimensional dual problem in Eq.~\eqref{eq:kl_dual} on the empirical loss distribution to obtain KL-robust upper and lower CVA bounds.

The KL radius controls how far the distribution can move: by Pinsker's inequality, the total-variation distance satisfies $\|Q-P\|_{\mathrm{TV}}\le 
\sqrt{D_{\mathrm{KL}}(Q\|P)/2}\le \sqrt{\varepsilon/2}$, which bounds the change in any event probability under $Q$ \citep{cover2006elements,csiszar2011information}.

\subsection{Calculation of $\varepsilon$ from market co-movements}\label{sec:eps_calib}
\paragraph{Step 1: estimate a stressed dependence target.}
Let $x_t$ denote daily log returns of an observed transition-risk market proxy and let $y_t$ denote daily changes in a credit-spread measure $\mathrm{CS}_t$, i.e., $y_t=\mathrm{CS}_t-\mathrm{CS}_{t-1}$. For each rolling window, we compute Kendall's rank correlation $\tau(x,y)$ and map it onto the Gaussian-copula latent-correlation scale via
\[
\rho = \sin\!\left(\frac{\pi}{2}\tau\right).
\]
This conversion is used only to place the empirical dependence target on a familiar correlation scale.

Because WWR is directional, we separate positive and negative dependence. We define one-sided components $\rho^{\text{long}}=\max(\rho,0)$ and $\rho^{\text{short}}=\max(-\rho,0)$, corresponding to exposure increasing when the transition-risk market factor rises or falls, respectively.
We focus on stressed co-movements by conditioning on a pre-defined stress regime (high realized volatility of the transition-risk market factor) and take a conservative tail quantile of $\rho^{\text{short}}$ to obtain the stressed empirical target $\rho_{\mathrm{target}}$.

\paragraph{Step 2: construct $\rho_{\mathrm{equiv}}(\varepsilon)$ and the budget $\varepsilon$.}
To give $\varepsilon$ a financial interpretation, we map the KL radius to the amount of dependence that a simple one-factor Gaussian-copula would generate on the equivalent exposure and default-time marginals.

First, for a grid of $\varepsilon$ values, we compute the KL-robust upper bound $\mathrm{CVA}^{\mathrm{upper}}(\varepsilon)$ using Eq.~\eqref{eq:kl_dual}. Because the headline reported object in Sections~\ref{sec:results} and~\ref{sec:nature} is scenario-relative ECVA rather than level CVA, the benchmark implementation performs the $\rho\leftrightarrow\varepsilon$ inversion on the same scenario-relative object; a raw scenario-level mapping is retained only as an internal diagnostic. Second, for each simulated market path $p$ we compress the entire exposure path into a scalar loss score
\[
Z_p=\sum_i DF(0,t_i)\,E_p^+(t_i)\,\Delta p_i,
\]
where $\Delta p_i=S(t_{i-1})-S(t_i)$ are the baseline interval default probabilities of the scenario under consideration. Let $o(1),\dots,o(P)$ denote the path indices sorted so that
$Z_{o(1)}\le \cdots \le Z_{o(P)}$.
The score $Z_p$ is therefore a deterministic path-ranking statistic computed from simulated exposures. To avoid confusion, we use a different notation for the auxiliary Gaussian draws that generate the copula uniforms. Two independent standard normals $G_{1,k},G_{2,k}\sim N(0,1)$ are used as
\[
U^{m}_k=\Phi(G_{1,k}),\qquad
U^{d}_k(\rho)=\Phi\!\left(-\rho G_{1,k}-\sqrt{1-\rho^2}\,G_{2,k}\right),
\]
where $\Phi$ is the standard normal cdf.
The market-path index is chosen by rank matching,
\[
p_k=o\!\left(\min\{\lfloor P\,U^{m}_k\rfloor+1,\;P\}\right),
\]
so larger $G_{1,k}$ selects a higher-score market path.
Default timing is generated from the baseline cumulative default masses
\[
F_i=\sum_{j=1}^{i}\Delta p_j,\qquad i=1,\dots,n,
\]
via inverse transform:
\[
I_k(\rho)=\min\{i:\,F_i\ge U^{d}_k(\rho)\}\quad \text{if }U^{d}_k(\rho)\le F_n,
\]
and otherwise no default occurs before maturity.
The resulting diagnostic loss is
\[
L_k(\rho)=
(1-R)\,DF(0,t_{I_k(\rho)})\,E^+_{p_k}(t_{I_k(\rho)}),
\]
with $L_k(\rho)=0$ when there is no default before maturity, and
\[
\mathrm{CVA}(\rho)=\frac{1}{N}\sum_{k=1}^{N}L_k(\rho).
\]
Applying the same diagnostic Gaussian-copula construction to scenario $s$ and to the reference scenario $ref$ yields $\mathrm{CVA}_s(\rho)$ and $\mathrm{CVA}_{ref}(\rho)$, and therefore
\[
\mathrm{ECVA}^{\rho}_s=\mathrm{CVA}_s(\rho)-\mathrm{CVA}_{ref}(\rho),
\qquad
\mathrm{ECVA}^{0}_s=\mathrm{CVA}_s(0)-\mathrm{CVA}_{ref}(0).
\]
For each $\varepsilon$ we already know $\mathrm{ECVA}^{\mathrm{upper}}_s(\varepsilon)$ and $\mathrm{ECVA}^{\mathrm{ind}}_s$ from Eqs.~\eqref{eq:ecva_ind}--\eqref{eq:wwr_did}. In the benchmark implementation we therefore define the scenario-relative add-ons
\[
A^{\mathrm{ECVA}}_{\varepsilon,s}
=
\mathrm{ECVA}^{\mathrm{upper}}_s(\varepsilon)-\mathrm{ECVA}^{\mathrm{ind}}_s,
\qquad
A^{\mathrm{ECVA}}_{\rho,s}
=
\mathrm{ECVA}^{\rho}_s-\mathrm{ECVA}^{0}_s.
\]
The equivalent model-implied correlation $\rho_{\mathrm{equiv}}(\varepsilon)$ used in the headline implementation is then obtained by
$A^{\mathrm{ECVA}}_{\varepsilon,s}\approx A^{\mathrm{ECVA}}_{\rho_{\mathrm{equiv}},s}$, implemented by monotone interpolation on the discrete $\rho$-grid. Finally, using the stressed target $\rho_{\mathrm{target}}$, we choose the dependence budget as
\[
\varepsilon^{\star}
=
\inf\Bigl\{\varepsilon\ge 0: \rho_{\mathrm{equiv}}(\varepsilon)\ge \rho_{\mathrm{target}}\Bigr\},
\]
implemented on a discrete monotone grid in $\rho$ and $\varepsilon$.

\section{Setup and results}\label{sec:results}
We take \emph{Current Policies} as the reference scenario and compare three NGFS Phase~V scenarios: \emph{Nationally Determined Contributions (NDCs)}, \emph{Net Zero 2050}, and \emph{Delayed Transition} \citep{NGFSPhaseVPortal,NGFSPhaseVUserGuide}.
For each scenario, we extract annual World series of GDP at purchasing power parity (NGFS variable \texttt{GDP|PPP}) and carbon prices (NGFS variable \texttt{Price|Carbon}), aggregating across NGFS models using the median. We then interpolate these series on the CVA time grid, construct the absolute multiplier as described in Section~\ref{sec:method}, and normalize it by the Current Policies multiplier to obtain the reference-relative hazard multiplier $m_s(t)$.
We build the discount curve from constant-maturity U.S. Treasury yields at standard maturities obtained from FRED \citep{FRED}, converting the quoted yields into discount factors under a continuous-compounding approximation.

The annual NGFS paths are well suited to forming long-horizon hazard multipliers, but they are too coarse to identify a daily dependence for the $\varepsilon$ calibration. We use raw WTI log returns (FRED series \texttt{DCOILWTICO}) together with changes in the ICE BofA US High Yield Option-Adjusted Spread (FRED series \texttt{BAMLH0A0HYM2}) \citep{FRED} to keep $\varepsilon$ tied to credit-market stress. 

\begin{table}[!htbp]
\centering
\caption{Benchmark summary.}
\label{tab:benchmark_spec}
\small
\begin{tabular}{p{0.43\linewidth} p{0.51\linewidth}}
\toprule
Item & Value \\
\midrule
As-of date & 2025-12-22 \\
Exposure object & 30-year payer-fixed interest-rate swap under HW1F\\
Notional & \$10,000,000 \\
Swap cashflow frequency & quarterly (4 per year) \\
Time step ($dt$) & 0.25 year \\
EPE Monte Carlo paths & 50,000 \\
Hull--White 1F mean reversion $a$ & 0.5 \\
Hull--White 1F volatility $\sigma$ & 0.01 \\
Recovery $R$ & 0.40 \\
Hazard calibration & CDS-par flat $\lambda_0$ at 5y  \\
NGFS database & Phase V (World); reference scenario $ref=$ Current Policies\\
Scenario-to-credit drivers & NGFS variables \texttt{GDP|PPP} and \texttt{Price|Carbon}\\
Translation coefficients & $\beta_{\mathrm{GDP}}=-0.6,\;\beta_{\mathrm{Carbon}}=0.15$\\
Dependence calibration  & Raw WTI log returns (\texttt{DCOILWTICO}) vs HY OAS changes (\texttt{BAMLH0A0HYM2})\\
OAS series &
 BAMLC0A0CM\\
OAS (pct) &
 0.8\\
 R &
 0.4 \\
 $\lambda_0$ &
 0.0133 \\
\bottomrule
\end{tabular}

\normalsize
\end{table}

Table~\ref{tab:benchmark_spec} summarizes the benchmark setting. The reduced-form NGFS-to-credit mapping in Section~\ref{sec:results} uses two elasticities, $(\beta_{\mathrm{GDP}},\beta_{\mathrm{CP}})$, to translate long-horizon GDP and carbon-price scenario objects into hazard multipliers.
Because NGFS GDP and carbon-price paths are scenario objects rather than historical realizations, they do not statistically identify these coefficients by regression. The baseline intensity $\lambda_0$ is calibrated by solving the CDS par-spread equation under a flat-intensity assumption. Table~\ref{tab:ngfs_multipliers} reports the baseline hazard calibration and selected NGFS hazard multipliers at representative horizons.

\begin{table}[!htbp]
\centering
\caption{Selected NGFS hazard multipliers relative to the reference scenario at representative horizons.}
\label{tab:ngfs_multipliers}
\begin{tabular}{lllll}
\toprule
Scenario & m(5y) & m(10y) & m(20y) & m(30y) \\
\midrule
Current Policies & 1.0000 & 1.0000 & 1.0000 & 1.0000 \\
Net Zero 2050 & 1.2048 & 1.2557 & 1.3855 & 1.4107 \\
Delayed transition & 1.0000 & 1.4812 & 1.7278 & 1.8742 \\
Nationally Determined Contributions (NDCs) & 1.0965 & 1.0978 & 1.1003 & 1.0944 \\
\bottomrule
\end{tabular}

\end{table}

\begin{figure}[!htbp]
\centering
\includegraphics[width=0.85\linewidth]{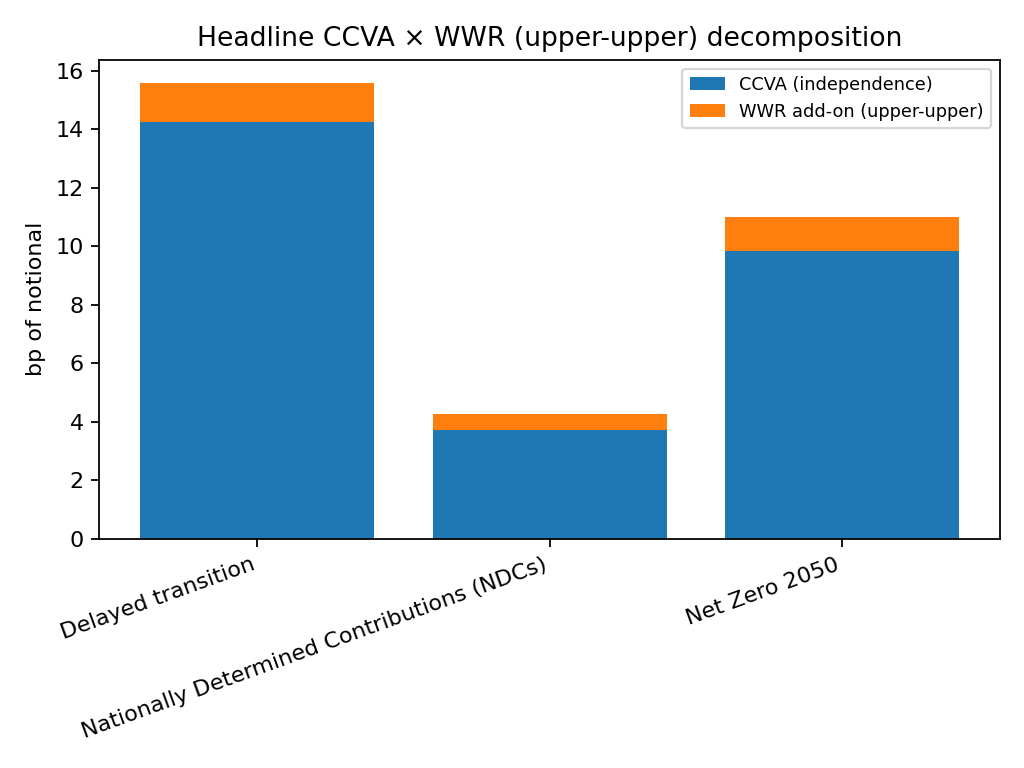}
\caption{Decomposition of headline CCVA into independence CCVA and the robust \WWRDelta.}
\label{fig:headline_decomp}
\end{figure}

Figure~\ref{fig:headline_decomp} reports the main decomposition in basis points of notional. The independence CCVA is positive across the transition scenarios considered, consistent with scenario-implied hazard multipliers that increase relative to the ``Current Policies'' reference.
The magnitude varies substantially across scenarios: Delayed transition produces the largest independence CCVA (14.57~bp), followed by Net Zero 2050 (9.73~bp) and NDCs (3.64~bp). The robust WWR is smaller but economically non-trivial, ranging from 0.56 to 1.61~bp (about 10--13\% of the total). Scenario-to-credit translation drives the first-order effect, while the robust WWR contributes a disciplined model-risk buffer around the independence benchmark.

\begin{figure}[!htbp]
\centering
\IfFileExists{figures/fig_step12_kl_marginal_distortion.png}{\includegraphics[width=0.88\linewidth]{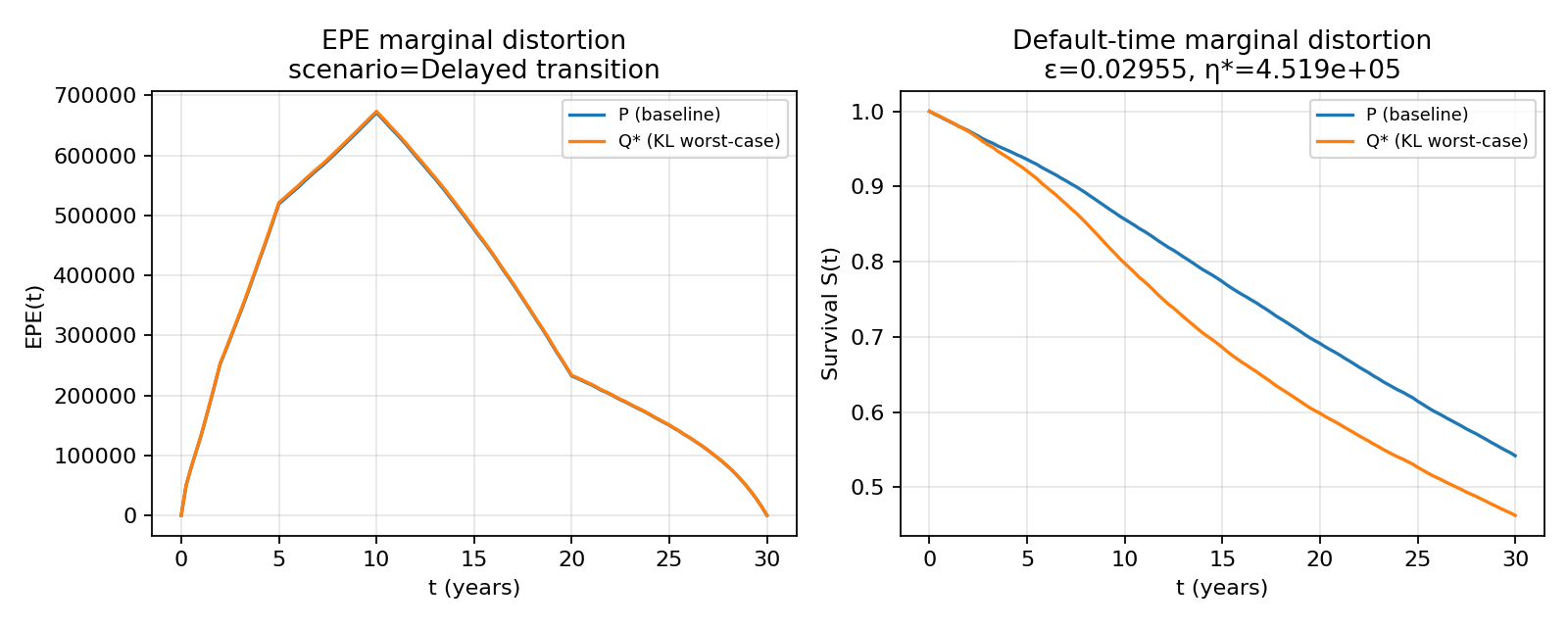}}{\fbox{\parbox{0.88\linewidth}{}}}
\caption{Marginal distortion induced by the KL-robust worst-case reweighting. The figure compares the baseline independence measure $P$ with the worst-case reweighted measure $Q^\star$ in terms of the default-time marginal distribution and the exposure marginal summarized by the EPE profile.}
\label{fig:kl_marginal_diag}
\end{figure}

Figure~\ref{fig:kl_marginal_diag} compares the baseline independence measure $P$ with the worst-case reweighted measure $Q^\star$ in terms of (i) the default-time marginal distribution and (ii) the exposure marginal summarized by EPE. Because the KL ball is applied to the \emph{joint} law of exposure paths and default times, both marginals can move in principle. In this example, however, the EPE profiles under $P$ and $Q^\star$ are nearly indistinguishable at plotting scale. The robust add-on is therefore generated mainly by reallocation of probability across default-time states rather than by a material deformation of the exposure marginal in this benchmark.

\begin{figure}[!htbp]
\centering
\IfFileExists{figures/fig_step12_eps_sweep_uplift.png}{\includegraphics[width=0.65\linewidth]{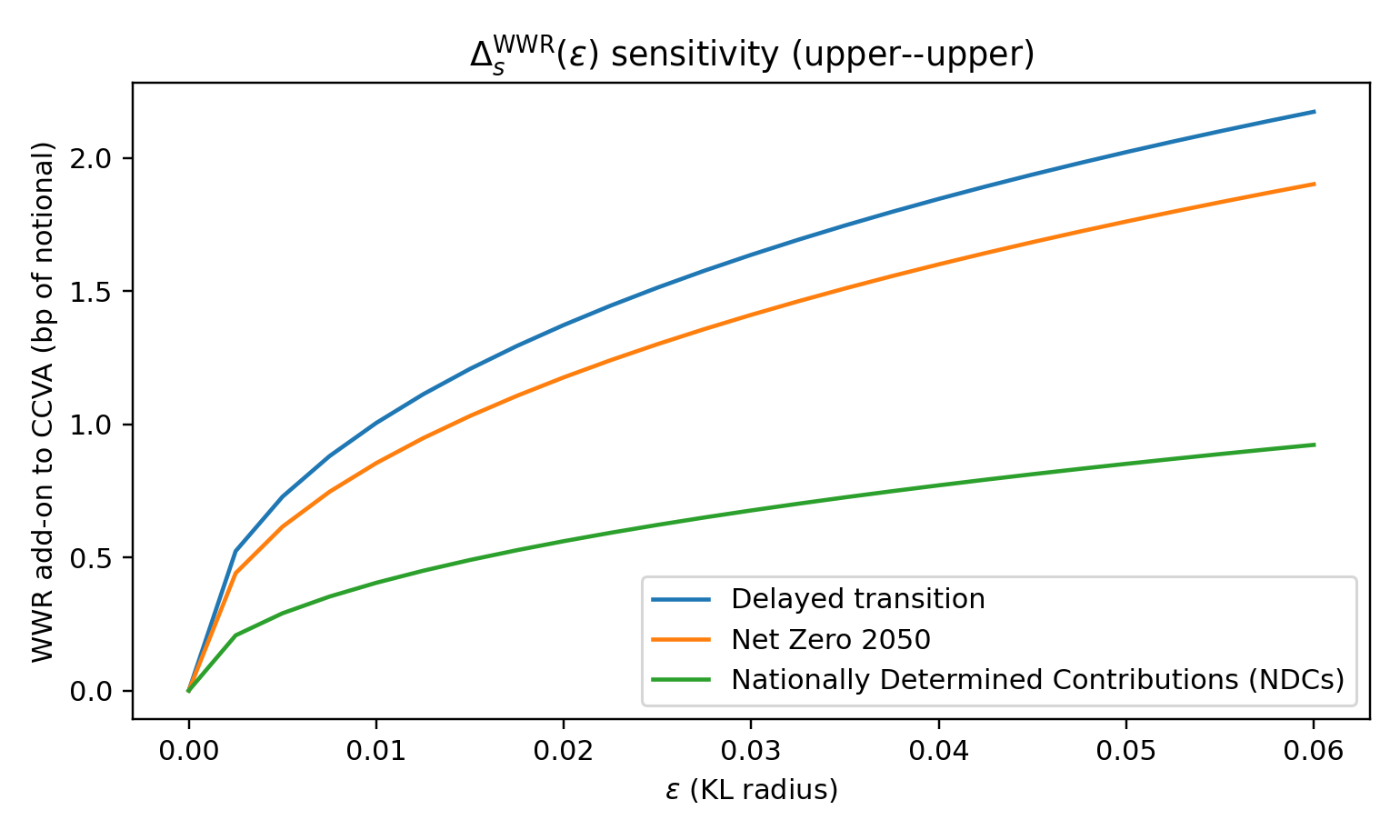}}{\fbox{\parbox{0.65\linewidth}{}}}
\caption{Sensitivity of the robust WWR to the KL radius $\varepsilon$.}
\label{fig:eps_sweep}
\end{figure}

Figure~\ref{fig:eps_sweep} reports the robust \WWRDelta as a function of the KL radius $\varepsilon$ for the headline configuration. Appendix~\ref{app:wti_swap} adds a commodity-swap illustration in which transition stress affects both the hazard curve and a WTI forward curve; there the same CCVA logic yields a credit/market/interaction decomposition.

\FloatBarrier

\section{Nature CVA}\label{sec:nature}
This section reuses the CVA framework and KL-robust WWR construction from Section~\ref{sec:method}. The nature framework introduces two additional ingredients: a biodiversity-specific policy-to-hazard mapping, with an optional two-stage entity mapping for region-specific physical shocks, and tail generators that quantify scenario-generation model risk.

\subsection{Construction of hazard multiplier: Reduced-form translation}\label{sec:policy_translation}
If a nature indicator $x_s(t)$ under scenario $s$ (e.g., biodiversity intactness) is available, we can map it into a hazard multiplier. We treat $x$ as a \emph{good-state} indicator (\emph{good is up}): larger $x$ means a healthier nature state.
If an upstream source defines an indicator with the opposite sign (\emph{bad is up}), we first apply a monotone transformation to convert it into a good-state form before constructing stress ratios. For indicators where \emph{lower is worse} (``bad is down''), define the scenario stress relative to a base year $t_0$ as
\begin{equation}
\mathrm{stress}_s(t) = \frac{x_s(t_0)}{x_s(t)}.
\end{equation}
Let ``ref'' denote a reference scenario and define the reference-relative stress ratio
\begin{equation}
\mathrm{SR}_s(t) = \frac{\mathrm{stress}_s(t)}{\mathrm{stress}_{\mathrm{ref}}(t)}.
\end{equation}
The \emph{policy} hazard multiplier is
\begin{equation}
m_s^{\mathrm{policy}}(t) = \mathrm{clip}\!\left(\mathrm{SR}_s(t)^{\beta},\ \underline{m},\ \overline{m}\right),
\end{equation}
where $\beta$ is a benchmark translation sensitivity parameter and $\underline{m},\overline{m}$ are floor/cap values for numerical stability.
Finally, the scenario intensity is given by
\begin{equation}
\lambda_s^{\mathrm{policy}}(t) = m_s^{\mathrm{policy}}(t)\,\lambda_0(t),
\end{equation}
where $\lambda_0(t)$ is the baseline intensity calibrated from market data, and we set $\beta=1$ in the benchmark. In an empirical implementation, $\beta$ would be estimated from issuer- or sector-level panels by regressing log changes in a credit proxy (for example CDS-implied hazard, OAS, EDF, or default-frequency proxy) on log changes in the constructed stress ratio, with issuer and time fixed effects and sectoral controls. Because matched biodiversity/credit panels are still sparse and heterogeneous, we treat $\beta$ here as a transparent governance parameter.

\subsection{Construction of hazard multiplier: Two-stage transmission layer}\label{sec:two_stage}
When the available scenario object is a regional physical variable rather than a direct biodiversity indicator, we can use a two-stage mapping from physical shock to firm cash-flow / operating-margin shock and then from those shocks to credit. Define the physical shock ratio
\[
\mathrm{cr}_s(t)=\frac{x_s(t)}{x_{\mathrm{ref}}(t)}
\]
and in order to represent partial dependence on the affected region, we can introduce an exposure share $\omega\in[0,1]$ and define an effective catch ratio
\[
\mathrm{cr}^{\mathrm{eff}}_s(t)=1+\omega\big(\mathrm{cr}_s(t)-1\big).
\]
We then map the physical ratio into a firm revenue ratio
\[
\mu_s(t)=\big(\mathrm{cr}^{\mathrm{eff}}_s(t)\big)^{\alpha_{\mathrm{catch}}}\,\pi_s(t)^{\alpha_{\mathrm{price}}},
\]
where $\pi_s(t)$ is an optional price-proxy term (e.g., fishmeal prices). The elasticity $\alpha_{\mathrm{price}}$ can be estimated jointly with $\alpha_{\mathrm{catch}}$ from a log-linear pass-through regression,
\[
\log\frac{\mathrm{Rev}_t}{\mathrm{Rev}_0}
=
\alpha_{\mathrm{catch}}\log\frac{Q_t}{Q_0}
+
\alpha_{\mathrm{price}}\log\frac{P_t}{P_0}
+\eta_t,
\]
using firm, segment, or peer-panel data, where $Q_t$ is a quantity proxy and $P_t$ is the chosen output-price proxy. In the following Austevoll example, we set $\pi_s(t)=1$ and $\alpha_{\mathrm{price}}=1$, because we do not impose an explicit future fishmeal-price scenario and want the benchmark to isolate quantity-driven nature stress.

Given a baseline cost share $c\in(0,1)$, we can also use the implied EBITDA-like ratio
\[
e_s(t)=\frac{\mu_s(t)-c}{1-c},
\]
and we compress that into a signed badness index
\[
u_s(t)=-\log e_s(t).
\]
Thus $e_s(t)<1$ (equivalently $u_s(t)>0$) means margin compression, while $e_s(t)>1$ means improvement. To keep the recovery adjustment conservative, we use the one-sided adverse component $u_s^+(t)=\max\{u_s(t),0\}$.
The credit translation is then
\[
m_s(t)=\mathrm{clip}\big(\exp(\beta_{\mathrm{PD}} u_s(t)),\, m_{\min},m_{\max}\big),\qquad
R_s(t)=\mathrm{clip}\big(R_0-k_R u_s^+(t),\, R_{\min},R_{\max}\big),
\]
and we set $\lambda_s(t)=m_s(t)\lambda_0$.
With richer segment or accounting data, $c$, $\omega$, and the elasticities can be replaced by firm-specific estimates or disclosures. A practical estimation route for $\beta_{\mathrm{PD}}$ is then a second-step panel regression of changes in a credit proxy---such as CDS spreads, bond OAS, EDF, or a mapped hazard rate---on the constructed badness index $u_{i,t}$ (or its horizon-$h$ counterpart), again with issuer and time fixed effects. In other words, the first step estimates the operating pass-through from physical shock to earnings pressure, and the second step estimates the credit sensitivity of that earnings-pressure index. In this paper we keep $\beta_{\mathrm{PD}}$ as a benchmark governance parameter.

\subsection{Tail generator}\label{sec:tail_generator}
Biodiversity and ecosystem dynamics are nonlinear and may exhibit regime shifts, path dependence, and stochasticity.
In this case, treating the deterministic policy scenario as the future path is not enough; the shape of the tail becomes a modeling choice.

The deterministic BES-SIM path provides the starting point \citep{kim2018bessim,pereira2024bessim}.
BES-SIM (\emph{Biodiversity and Ecosystem Services Scenario-based Intercomparison of Models}) is a scenario-based intercomparison framework in which harmonized land-use and climate pathways are fed into multiple ecological models. BES-SIM produces an ensemble of model-dependent time series for biodiversity and ecosystem services.

For each tail provider, we first construct a panel of yearly pathwise stress series $g_i(t)$ so that $g_i(t_0)=1$.
We then convert those provider paths into tail factors by year-wise normalization:
\[
f_i(t)=\frac{g_i(t)}{\operatorname{median}_j g_j(t)}.
\]
The one-sided construction replaces $f_i(t)$ by $\max\{1,f_i(t)\}$ so that only adverse tail deformations are considered. The hybrid stress is then
\[
\mathrm{SR}^{\mathrm{hyb}}_{s}(t)=\frac{\mathrm{stress}_s(t)\cdot f_i(t)}{\mathrm{stress}_{\mathrm{ref}}(t)},
\]
and the corresponding hybrid multiplier is
\begin{equation}\label{eq:hyb_multiplier}
m_{s,i}(t)=\mathrm{clip}\!\left(\mathrm{SR}^{\mathrm{hyb}}_{s}(t)^{\beta},\ \underline{m},\overline{m}\right).
\end{equation}
This makes the policy$\times$tail construction explicit: the deterministic policy path anchors the center of the distribution, and the tail generator deforms that path multiplicatively year by year.

In this paper, we construct tail-generating paths from two distinct provider families that capture different sources of uncertainty.
For MadingleyR\citep{harfoot2014madingley,hoeks2021madingleyr}, provider paths are obtained by repeatedly simulating the same mechanistic ecosystem model under different random seeds, thereby capturing stochastic variability.
For ISIMIP, provider paths are drawn from the publicly available ISIMIP3b archive \citep{ISIMIPProtocol3}. The broader ISIMIP framework provides harmonized multi-model impact simulations, and in our implementation the distributed CLASSIC \texttt{cveg} provider set is used to proxy model uncertainty \citep{warszawski2014isimip}.
Writing $g_{m,k}(t)$ for the annual path produced by ecosystem model $m$ and member $k$ under a common ISIMIP scenario, we use the distributed set $\{g_{m,k}(t)\}_{m,k}$ as the source of provider paths.
In the benchmark specification, we focus on the annual CLASSIC \texttt{cveg} subset—where CLASSIC denotes one of the terrestrial ecosystem models included in ISIMIP and \texttt{cveg} denotes its vegetation carbon output.

\subsection{Data sources and scenario objects}\label{sec:data_case}
For BES-SIM, we select the \emph{Global} region and the \emph{PREDICTS} model. Specifically, we take the published PREDICTS intactness series, interpolate it to annual values on 2025--2054, and use that annual path as the deterministic policy object that enters the benchmark hazard multiplier. The reference scenario is \texttt{SSP1xRCP2.6} and the stressed scenarios are \texttt{SSP3xRCP6.0} and \texttt{SSP5xRCP8.5}. SSP denotes a \emph{Shared Socioeconomic Pathway}, and RCP denotes a \emph{Representative Concentration Pathway}. The number after ``SSP'' identifies the socioeconomic storyline rather than a physical forcing level: SSP1 is a sustainability-oriented pathway, SSP3 is a regional-rivalry / fragmented-development pathway, and SSP5 is a fossil-fueled development / high-energy-demand pathway \citep{kim2018bessim,pereira2024bessim}. By contrast, the number after ``RCP'' (e.g., 2.6, 6.0, 8.5) is the approximate radiative forcing in 2100 (W/m$^2$) and is widely used as a compact marker of scenario severity. Table~\ref{tab:benchmark_spec_nature} summarizes the benchmark configuration.

\begin{table}[!htbp]
\centering
\caption{Benchmark configuration (Nature CVA).}
\label{tab:benchmark_spec_nature}
\small
\setlength{\tabcolsep}{4pt}
\renewcommand{\arraystretch}{1.05}
\begin{tabular}{p{0.24\textwidth}p{0.72\textwidth}}
\toprule
Item & Setting \\
\midrule
Horizon & 2025--2054 (annual grid) \\
Exposure object & Same as in the climate benchmark \\
Discounting & Same as in the climate benchmark \\
Credit baseline & Same as in the climate benchmark \\
Policy translation & $m^{\mathrm{policy}}_s(t)=\mathrm{clip}(\mathrm{SR}_s(t)^{\beta},[0.05,20])$ with $\beta=1$ \\
WWR robustness & KL radius $\varepsilon=0.003$ (default) and calibrated $\varepsilon=0.01948$ (Table~\ref{tab:wwr_rep}); loss-sample MC size $50{,}000$ \\
\bottomrule
\end{tabular}
\end{table}

\FloatBarrier

\begin{figure}[!htbp]
\centering
\includegraphics[width=0.88\linewidth]{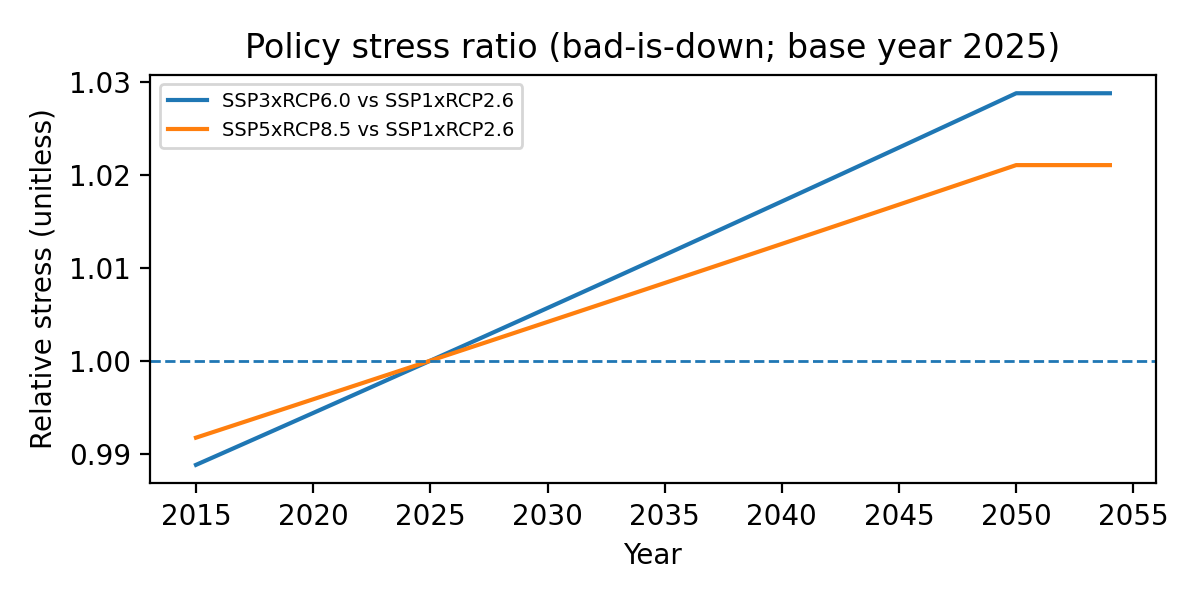}
\caption{Deterministic policy stress ratios $\mathrm{SR}_s(t)$ constructed from BES-SIM PREDICTS intactness.}
\label{fig:policy_stress}
\end{figure}

Figure~\ref{fig:policy_stress} shows the deterministic stress ratio $\mathrm{SR}_s(t)$ constructed from the BES-SIM PREDICTS intactness paths (annualized by linear interpolation between available endpoints).
This policy-only translation produces NCVA of about 0.75--1.02~bp for SSP5--RCP8.5 and SSP3--RCP6.0, respectively (Table~\ref{tab:ncva_summary}).
Because the policy translation uses a \emph{global-average} intactness series, scenario differences are mechanically attenuated.
The resulting policy-only NCVA should therefore be read as a conservative lower-bound magnitude for more granular sector and region-specific mappings.

\FloatBarrier

\subsection{Tail generator comparison: MadingleyR vs ISIMIP}\label{sec:tail_compare}
We next implement the hybrid policy$\times$tail construction in Eq.~\eqref{eq:hyb_multiplier}: the deterministic BES-SIM policy stress is held fixed, while alternative ecosystem-model ensembles provide the tail factors $f_i(t)$ that deform the NCVA distribution.
We use MadingleyR and ISIMIP and apply the same normalization to both generators: year by year we divide each provider path by the provider median to obtain the two-sided factor $f_i(t)$, and we replace $f_i(t)$ by $\max\{1,f_i(t)\}$ for the one-sided conservative variant.

\begin{figure}[!htbp]
\centering
\begin{subfigure}{0.48\linewidth}
\centering
\includegraphics[width=\linewidth]{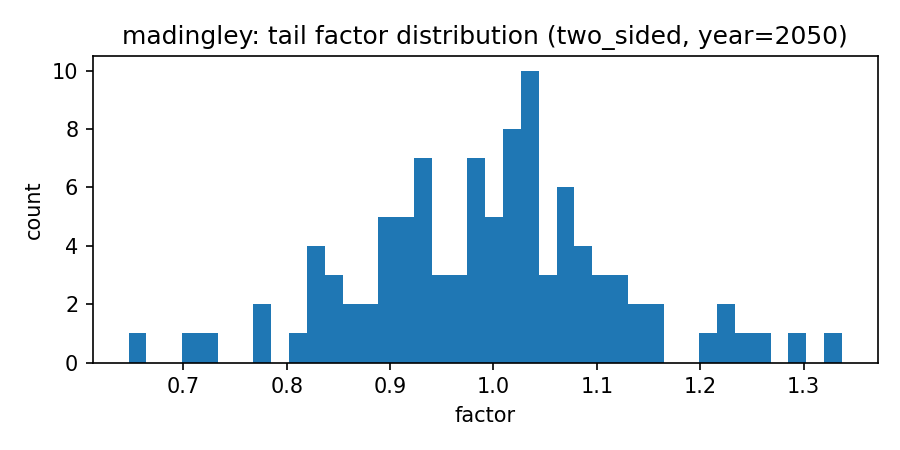}
\caption{MadingleyR (seed ensemble).}
\end{subfigure}\hfill
\begin{subfigure}{0.48\linewidth}
\centering
\includegraphics[width=\linewidth]{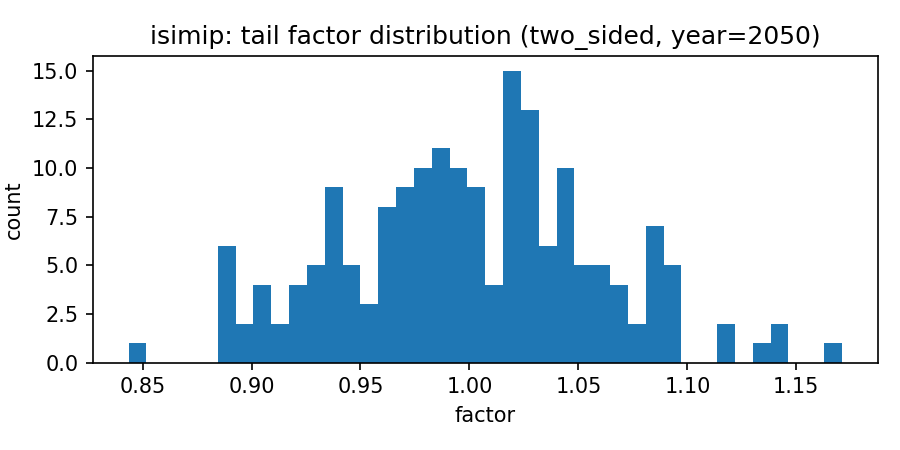}
\caption{ISIMIP3b biomes (CLASSIC \texttt{cveg}).}
\end{subfigure}
\caption{Tail-factor distribution at year 2050 (two-sided mode). In the benchmark, MadingleyR exhibits a substantially heavier right tail than ISIMIP.}
\label{fig:tail_factors}
\end{figure}

Figure~\ref{fig:tail_factors} first shows the year-2050 tail-factor distribution under the two providers. In this benchmark, MadingleyR exhibits a visibly heavier right tail than ISIMIP. A plausible reason is that MadingleyR retains within-model seed stochasticity and nonlinear ecological propagation, whereas the ISIMIP benchmark uses annual global CLASSIC \texttt{cveg} paths that are smoother before median normalization.

\FloatBarrier

\begin{figure}[!htbp]
\centering
\begin{subfigure}{0.48\linewidth}
\centering
\includegraphics[width=\linewidth]{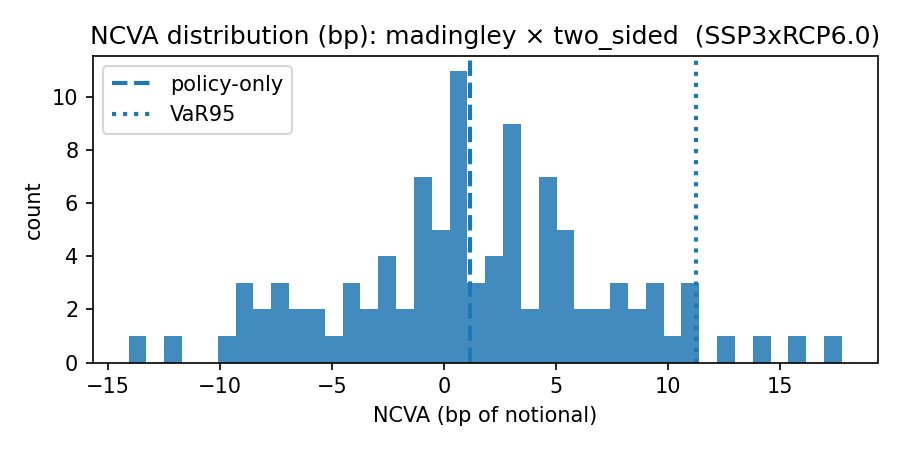}
\caption{SSP3--RCP6.0 (two-sided).}
\end{subfigure}\hfill
\begin{subfigure}{0.48\linewidth}
\centering
\includegraphics[width=\linewidth]{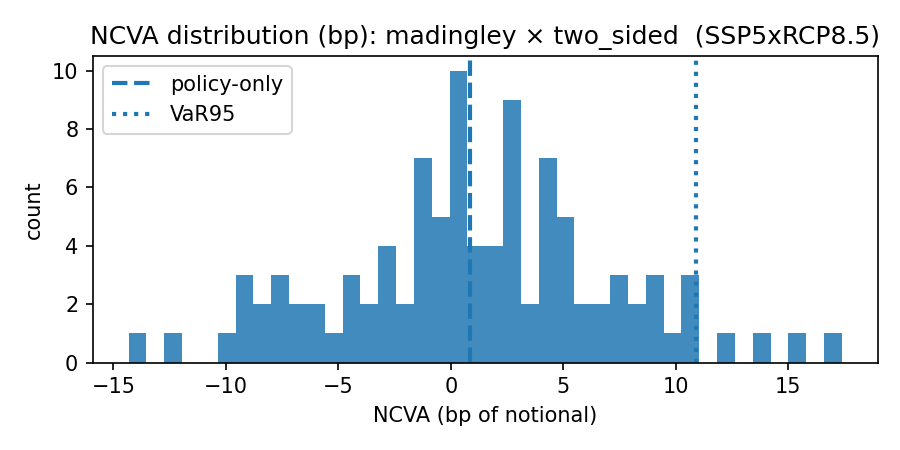}
\caption{SSP5--RCP8.5 (two-sided).}
\end{subfigure}
\caption{NCVA distributions under hybrid policy$\times$tail scenarios (two-sided mode).}
\label{fig:ncva_compare}
\end{figure}

Figure~\ref{fig:ncva_compare} reports the resulting NCVA distributions under the two tail generators, and Table~\ref{tab:ncva_summary} summarizes policy-only NCVA together with the hybrid distributional statistics (median/mean/VaR/ES).
Under the global-average intactness input, policy-only NCVA remains modest, but the tail behavior is much more model-dependent: MadingleyR produces materially heavier tails than ISIMIP.
This indicates that scenario-generation uncertainty is quantitatively important even when the deterministic policy path is held fixed. Table~\ref{tab:wwr_rep} reports independence NCVA, the KL-robust WWR, their difference ($\Delta\mathrm{WWR}$), and total NCVA.

\FloatBarrier

\begin{table}[!htbp]
\centering
\caption{NCVA distribution summary (bp of notional) under hybrid policy$\times$tail scenarios. ``Policy-only'' is the deterministic NCVA from the BES-SIM translation layer; other statistics are computed over the tail-generator path ensemble.}
\label{tab:ncva_summary}
\small
\resizebox{\textwidth}{!}{%
\begin{tabular}{ll l r r r r r r r r r}
\toprule
Scenario & Provider & Mode & $N$ & Policy-only & Median & Mean & VaR95 & ES95 & VaR99 & ES99 & $P(\mathrm{NCVA}<0)$ \\
\midrule
SSP3--RCP6.0 & ISIMIP (Biomes) & one-sided & 180 & 1.016 & 1.305 & 1.690 & 3.483 & 4.330 & 5.130 & 5.232 & 0.000 \\
SSP3--RCP6.0 & ISIMIP (Biomes) & two-sided & 180 & 1.016 & 1.054 & 1.016 & 3.483 & 4.330 & 5.130 & 5.232 & 0.283 \\
SSP3--RCP6.0 & MadingleyR & one-sided & 100 & 1.016 & 2.967 & 4.218 & 13.022 & 14.453 & 14.439 & 16.486 & 0.000 \\
SSP3--RCP6.0 & MadingleyR & two-sided & 100 & 1.016 & 1.566 & 1.716 & 12.951 & 14.447 & 14.412 & 16.486 & 0.400 \\
SSP5--RCP8.5 & ISIMIP (Biomes) & one-sided & 180 & 0.747 & 1.035 & 1.418 & 3.202 & 4.046 & 4.848 & 4.945 & 0.000 \\
SSP5--RCP8.5 & ISIMIP (Biomes) & two-sided & 180 & 0.747 & 0.784 & 0.747 & 3.202 & 4.046 & 4.848 & 4.945 & 0.339 \\
SSP5--RCP8.5 & MadingleyR & one-sided & 100 & 0.747 & 2.691 & 3.937 & 12.710 & 14.135 & 14.122 & 16.160 & 0.000 \\
SSP5--RCP8.5 & MadingleyR & two-sided & 100 & 0.747 & 1.295 & 1.444 & 12.639 & 14.129 & 14.095 & 16.160 & 0.410 \\
\bottomrule
\end{tabular}%
}
\end{table}

\FloatBarrier

\begin{table}[!htbp]
\centering
\caption{KL-robust WWR decomposition (two-sided). Values are in bp of notional for paths closest to NCVA quantiles (median / VaR95 / VaR99).}
\label{tab:wwr_rep}
\small
\resizebox{\textwidth}{!}{%
\begin{tabular}{llrrrrrr}
\toprule
scenario & tail source & quantile & NCVA$^{\mathrm{ind}}$ (bp) & WWR add-on$_s$ (bp) & WWR add-on$_{ref}$ (bp) & $\Delta$WWR (bp) & NCVA$^{\mathrm{upper}}$ (bp) \\
\midrule
SSP3--RCP6.0 & ISIMIP & 0.50 & 1.0528 & 285.863 & 287.136 & -1.2733 & -0.2205 \\
SSP3--RCP6.0 & ISIMIP & 0.95 & 3.4799 & 283.980 & 287.136 & -3.1561 & 0.3238 \\
SSP3--RCP6.0 & ISIMIP & 0.99 & 5.1107 & 283.148 & 287.136 & -3.9887 & 1.1220 \\
SSP3--RCP6.0 & MadingleyR & 0.50 & 1.5016 & 287.164 & 287.136 & 0.0274 & 1.5291 \\
SSP3--RCP6.0 & MadingleyR & 0.95 & 12.9443 & 277.664 & 287.136 & -9.4723 & 3.4721 \\
SSP3--RCP6.0 & MadingleyR & 0.99 & 14.3908 & 274.195 & 287.136 & -12.9415 & 1.4493 \\
SSP5--RCP8.5 & ISIMIP & 0.50 & 0.7855 & 286.433 & 287.136 & -0.7036 & 0.0819 \\
SSP5--RCP8.5 & ISIMIP & 0.95 & 3.1989 & 284.403 & 287.136 & -2.7330 & 0.4659 \\
SSP5--RCP8.5 & ISIMIP & 0.99 & 4.8293 & 283.352 & 287.136 & -3.7848 & 1.0444 \\
SSP5--RCP8.5 & MadingleyR & 0.50 & 1.3586 & 286.173 & 287.136 & -0.9632 & 0.3954 \\
SSP5--RCP8.5 & MadingleyR & 0.95 & 12.6330 & 278.883 & 287.136 & -8.2529 & 4.3801 \\
SSP5--RCP8.5 & MadingleyR & 0.99 & 14.0740 & 274.204 & 287.136 & -12.9327 & 1.1412 \\
\bottomrule
\end{tabular}%
}
\end{table}

\FloatBarrier

\subsection{Example: Peru-exposed seafood counterparty}\label{sec:case}\label{sec:entity_ext}
This subsection considers the two-stage transmission layer in Section~\ref{sec:two_stage} by tracing how climate-driven changes in Peru's marine ecosystem propagate into firm-level credit inputs and ultimately into nature CVA.

Our physical inputs are drawn from the FishMIP marine ecosystem intercomparison, as distributed through the ISIMIP archive. FishMIP is the fisheries and marine-ecosystem intercomparison project within ISIMIP, and BOATS is the marine ecosystem model that maps climate and ocean conditions into projected fish catch. Because the physical object is generated by emissions/forcing scenarios passed through climate models and then through a marine-ecosystem model, this case study can also be read as a \emph{climate-to-nature transmission} example within the broader Environmental CVA architecture rather than as a purely stand-alone nature shock.

We consider two alternative future pathways for greenhouse-gas emissions and radiative forcing, SSP126 and SSP585, which represent lower- and higher-forcing scenarios, respectively. These pathways are translated into climate and ocean time series by Earth-system models; in our benchmark illustration, we use \texttt{gfdl-esm4} and \texttt{ipsl-cm6a-lr}. These two models provide alternative physically consistent realizations of the same forcing pathways and therefore generate different climate and ocean trajectories under identical scenarios, providing a measure of climate-model uncertainty. The resulting climate-driven inputs are then fed into BOATS to produce projected catch paths.

We aggregate BOATS output over Peru's Exclusive Economic Zone (EEZ) using the Peru EEZ polygon constructed from the Marine Regions World EEZ dataset \citep{vliz2023eez}. For each climate-model forcing, we construct a catch-ratio path by dividing projected Peru EEZ catch under SSP585 by the corresponding projected catch under SSP126. This ratio measures the physical shock under the higher-forcing scenario relative to the lower-forcing benchmark.

\begin{figure}[!htbp]
\centering
\includegraphics[width=0.80\linewidth]{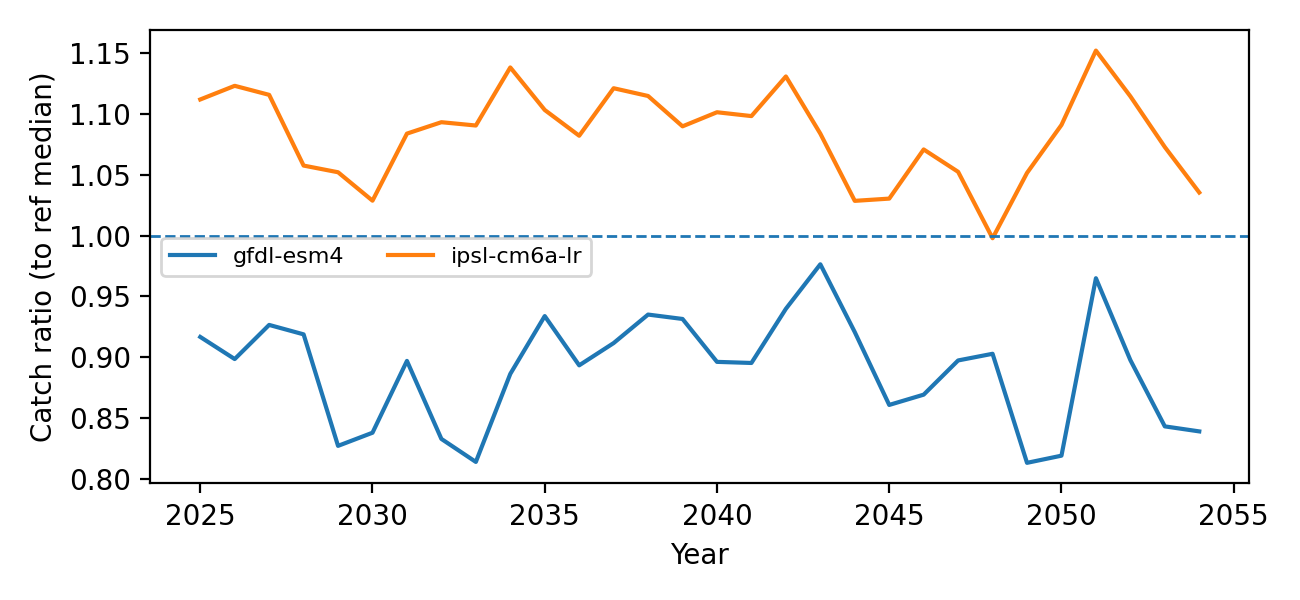}
\caption{Peru EEZ total catch ratio (SSP585 relative to SSP126) from BOATS, for two alternative climate-forcing members.}
\label{fig:austevoll_catch}
\end{figure}

Figure~\ref{fig:austevoll_catch} reports these catch-ratio paths for the two alternative climate-model forcings. Values above one indicate higher projected catch under SSP585 than under SSP126, while values below one indicate lower projected catch under the higher-forcing scenario. The two paths differ not only in magnitude but also in direction, reflecting physical-model uncertainty arising from alternative climate inputs rather than a precise predictive interval.

\FloatBarrier

In the second stage, we map the physical catch shock into firm-level credit quantities. Climate-driven changes in Peru catch volumes affect the Peru-exposed portion of the counterparty's business, which in turn shift operating performance and the credit-risk objects used in the CVA calculation.

Table~\ref{tab:austevoll_inputs} reports the benchmark parameterization used to translate the Peru EEZ shock into credit objects. In particular, the benchmark sets $\omega=1$, $c=0.65$, and $\beta_{\mathrm{PD}}=2.0$, with a two-sided hazard response and a one-sided recovery response. Table~\ref{tab:austevoll_case} summarizes the implied hazard multipliers and NCVA for the exposure and shows that localized physical shocks can generate economically non-trivial NCVA shifts. The construction is already compatible with an integrated climate--nature CVA. The upstream scenario is the same emissions/forcing scenario used in climate analysis; what changes is that the scenario is propagated through an additional biophysical block before entering hazard and recovery. A unified Environmental CVA could therefore combine direct climate-to-credit effects with indirect climate-to-nature-to-credit effects, provided the joint simulation is designed to avoid double counting.
\begin{table}[!htbp]
\centering
\caption{Case-study benchmark inputs. These are the explicit transmission parameters used to map Peru EEZ catch shocks into firm-level hazard and recovery objects.}
\label{tab:austevoll_inputs}
\small
\begin{tabular}{p{0.34\linewidth}p{0.58\linewidth}}
\toprule
Item & Benchmark setting \\
\midrule
Exposure& 30-year swap \\
Physical shock object & Peru EEZ BOATS total catch (\texttt{tc}), \texttt{SSP585} relative to \texttt{SSP126} \\
Discounting convention & Same as the climate and global nature benchmarks \\
Peru dependence share $\omega$ & 1.00 \\
Catch elasticity $\alpha_{\mathrm{catch}}$ & 1.00 \\
Price elasticity $\alpha_{\mathrm{price}}$ & 1.00 \\
Cost share $c$ & 0.65 \\
Baseline credit state & $\lambda_0=0.0155$, $R_0=0.40$ \\
Credit mapping & $\beta_{\mathrm{PD}}=2.0$, hazard clip $[0.2,8.0]$ \\
Recovery mapping & $k_R=0.05$, recovery bounds $[0.05,0.60]$, one-sided \\
\bottomrule
\end{tabular}
\end{table}

\FloatBarrier

\begin{table}[!htbp]
\centering
\caption{Case study using Peru EEZ fish-catch shocks from ISIMIP3b/FishMIP BOATS (total catch \texttt{tc}) under \texttt{SSP585} (SSP5--8.5) relative to \texttt{SSP126} (SSP1--2.6). Values are in bp of notional.}
\label{tab:austevoll_case}
\small
\begin{tabular}{l r r r r r r}
\toprule
Member & ${\mathrm{catch\ ratio}}$ & $[\min,\max]$ & $\overline{m}$ & $m(2054)$ & CVA (bp) & NCVA (bp)\\
\midrule
gfdl-esm4 & 0.890 & [0.814, 0.976] & 2.404 & 3.417 & 61.027 & 32.247\\
ipsl-cm6a-lr & 1.084 & [0.998, 1.152] & 0.669 & 0.825 & 19.965 & -8.815\\
\midrule
ref (median) & 1.000 & [1.000, 1.000] & 1.000 & 1.000 & 28.780 & 0.000\\
\bottomrule
\end{tabular}
\end{table}

\FloatBarrier

\section{Discussion}\label{sec:discussion}
In the climate framework, the dominant effect comes from the scenario-to-credit translation layer. Using NGFS scenarios, independence CCVA ranges from 3.64~bp (NDCs) to 14.57~bp (Delayed Transition) of notional in the headline configuration, while the KL-robust WWR add-on is smaller but still economically relevant. The transparent translation layer produces the first-order effect, and the robust WWR overlay provides an auditable second-order model-risk buffer. The nature CVA benchmark highlights that the distribution of NCVA is highly sensitive to the tail generator. Holding the deterministic BES-SIM policy path fixed, alternative ecosystem generators (ISIMIP versus MadingleyR) induce materially different tail behavior.
For nature risk, scenario generation is a quantitatively important source of model uncertainty. The Peru case study also suggests that the climate and nature blocks need not remain separate: when emissions scenarios are propagated through climate and ecosystem models before entering credit, the result is naturally interpreted as an integrated environmental transmission channel.

Our framework is reduced-form, which is both a strength and a limitation. It is a strength because each layer is explicit: discounting and exposure, credit-curve construction, scenario-to-credit translation, and dependence uncertainty. It is a limitation because several layers are still approximations. The translation elasticities are governance parameters rather than statistically identified structural coefficients, and the exposure engine omits collateral, netting, optionality, and multi-curve features.

\section{Conclusion}\label{sec:conclusion}
This paper develops an environmental CVA framework that treats climate risk and nature risk in a unified way. In the climate framework, NGFS scenarios are translated into hazard multipliers, while KL-robust bounds provide a conservative WWR buffer.
In the nature framework, the same scheme produces a policy-only NCVA benchmark from BES-SIM, and the policy$\times$tail construction shows that alternative ecosystem generators can produce materially different NCVA tails. The Peru illustration further shows how an emissions scenario can be carried through climate and ecosystem blocks before entering credit, pointing toward an integrated climate--nature CVA. The tail generator is itself part of the problem and should be treated as explicit model risk. Future research can include extension to multi-curve, collateralized, and portfolio-level settings, joint climate--nature simulation to avoid double counting, and estimation of sector- and region-specific translation elasticities.

\section*{Declaration of AI-assisted technologies}
During the preparation of this manuscript, the author used generative AI and AI-assisted tools to assist with manuscript drafting, language refinement, code drafting and debugging, and discussion of the presentation of the research. The author directed the project, reviewed, revised, and verified the manuscript and code, and takes full responsibility for the final manuscript.


\appendix

\section{Extension to WTI swap}\label{app:wti_swap}
The CCVA results in Section~\ref{sec:results} isolate the \emph{credit channel}: NGFS scenarios shift hazard curves with fixed exposures. However, for many commodity derivatives---especially energy products---a transition scenario may plausibly affect both (i) counterparty credit and (ii) the underlying commodity forward curve.

\begin{figure}[!htbp]
\centering
\includegraphics[width=0.78\linewidth]{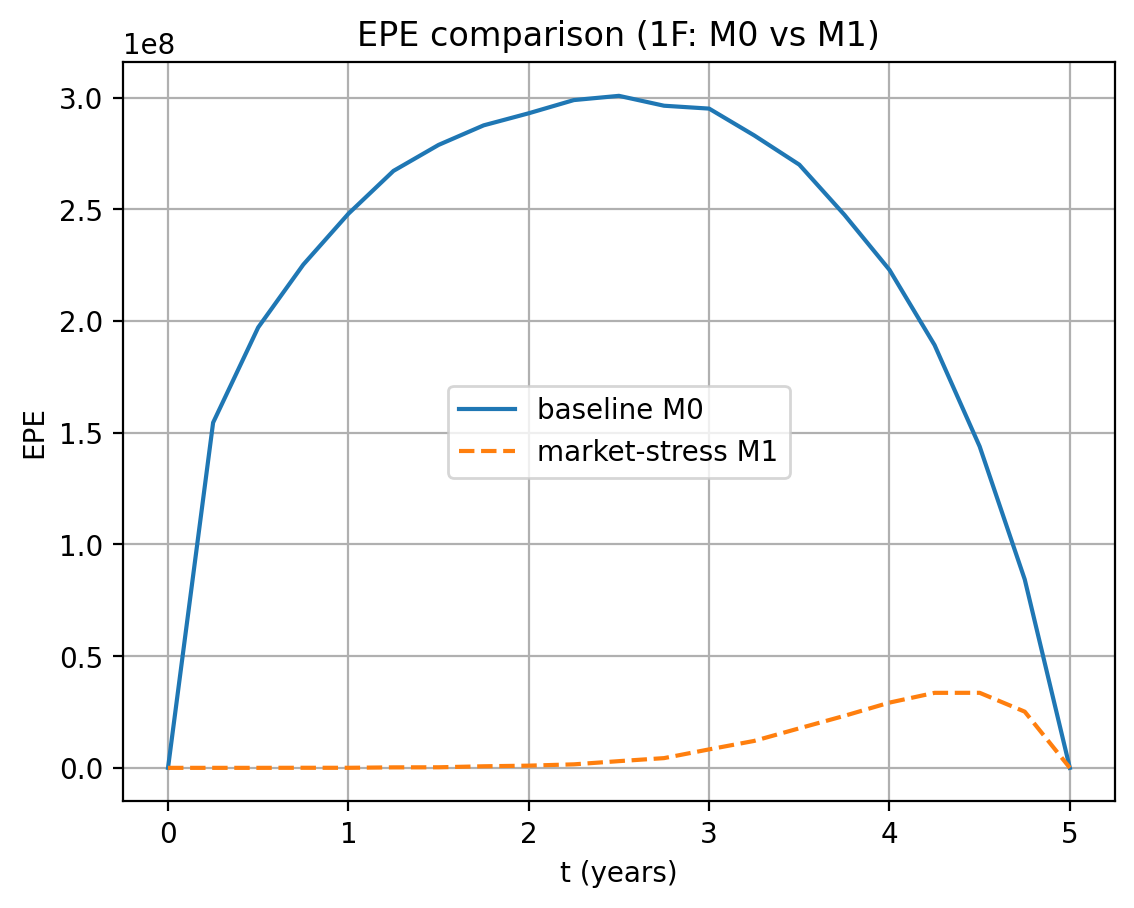}
\caption{WTI swap EPE under the one-factor Samuelson-type exposure model: baseline market state $M_0$ versus the market-stress state $M_1$ (15\% level-down forward curve).}
\label{fig:wti_epe_mkt}
\end{figure}

Let $\lambda_0$ denote the reference hazard term structure (Current Policies) and $\lambda_1$ the transition-stressed hazard term structure (Net Zero~2050) produced by the scenario-to-credit multiplier.
Let $M_0$ denote the baseline market state (WTI forward curve) and let $M_1$ denote a market-stress state.
Define the four ``corner'' CVAs by
\begin{equation}
\mathrm{CVA}_{ij}:=\mathrm{CVA}(\lambda_i, M_j),\qquad i,j\in\{0,1\}.
\end{equation}
Each corner uses the same independence discretization as Eq.~\eqref{eq:cva_discrete}, with the survival curve determined by $\lambda_i$ and the exposure distribution determined by the market state $M_j$:
\[
\mathrm{CVA}_{ij}
=(1-R)\sum_{k=1}^{n} DF(0,t_k)\,\mathrm{EPE}_{M_j}(t_k)\,\Big(S_{\lambda_i}(t_{k-1})-S_{\lambda_i}(t_k)\Big),
\]
where $\mathrm{EPE}_{M_j}(t_k)=\mathbb{E}[\max(V_{M_j}(t_k),0)]$ is produced by the WTI exposure simulation under market state $M_j$ and $S_{\lambda_i}$ is the survival curve implied by the hazard term structure $\lambda_i$.
Consequently, the joint commodity CCVA between the reference corner $(\lambda_0,M_0)$ and the stressed corner $(\lambda_1,M_1)$ can be decomposed as
\begin{equation}
\mathrm{CVA}_{11}-\mathrm{CVA}_{00}
=\underbrace{(\mathrm{CVA}_{10}-\mathrm{CVA}_{00})}_{\Delta\mathrm{CVA}_{\text{credit}}}
+\underbrace{(\mathrm{CVA}_{01}-\mathrm{CVA}_{00})}_{\Delta\mathrm{CVA}_{\text{market}}}
+\underbrace{(\mathrm{CVA}_{11}-\mathrm{CVA}_{10}-\mathrm{CVA}_{01}+\mathrm{CVA}_{00})}_{\Delta\mathrm{CVA}_{\text{interaction}}}.
\label{eq:credit_market_interaction}
\end{equation}
We implement a stylized 5-year WTI swap and simulate WTI futures under a one-factor model with exponentially damped maturity-dependent volatility, $\sigma(T,t)=\sigma_0 e^{-\kappa (T-t)}$, as a reduced-form Samuelson-type specification motivated by the maturity effect in futures markets \citep{Samuelson1965}.
To avoid mixing NGFS physical-measure scenario paths with risk-neutral pricing dynamics, we do not map NGFS oil-price paths directly into futures dynamics; instead, we run a stress test in which $M_1$ is a 15\% downward level shift of the initial WTI forward curve.
Figure~\ref{fig:wti_epe_mkt} shows that the market stress compresses EPE and shifts exposure weight toward the back of the horizon.

\begin{table}[!htbp]
\centering
\caption{WTI swap market-channel illustration (USD mm). Panel A reports the four ``corner'' CVAs $\mathrm{CVA}_{ij}=\mathrm{CVA}(\lambda_i,M_j)$, with $\lambda_0$ = Current Policies hazard, $\lambda_1$ = Net Zero~2050 hazard, $M_0$ = baseline WTI forward curve, and $M_1$ = a 15\% level-down stressed curve. Panel B reports the decomposition in Eq.~\eqref{eq:credit_market_interaction}; ``share'' is $|\Delta\mathrm{CVA}_{\text{interaction}}|/|\mathrm{CVA}_{11}-\mathrm{CVA}_{00}|$.}
\label{tab:wti_decomp_1f}
\begin{tabular}{lcc}
\toprule
Panel A: CVA (USD mm) & $M0$ & $M1$ \\
\midrule
Credit $\lambda_0$ (Current Policies) & 15.13 & 0.59 \\
Credit $\lambda_1$ (Net Zero 2050)    & 17.78 & 0.73 \\
\bottomrule
\end{tabular}

\vspace{0.8em}

\begin{tabular}{lrrrrr}
\toprule
Panel B: decomposition (USD mm) & $\Delta\mathrm{CVA}_{\text{credit}}$ & $\Delta\mathrm{CVA}_{\text{market}}$ & $\Delta\mathrm{CVA}_{\text{interaction}}$ & $\Delta\mathrm{CVA}$ & share \\
\midrule
1F+Samuelson & 2.64 & -14.55 & -2.50 & -14.41 & 17.4\% \\
\bottomrule
\end{tabular}

\end{table}

\FloatBarrier

\begin{figure}[!htbp]
\centering
\includegraphics[width=0.78\linewidth]{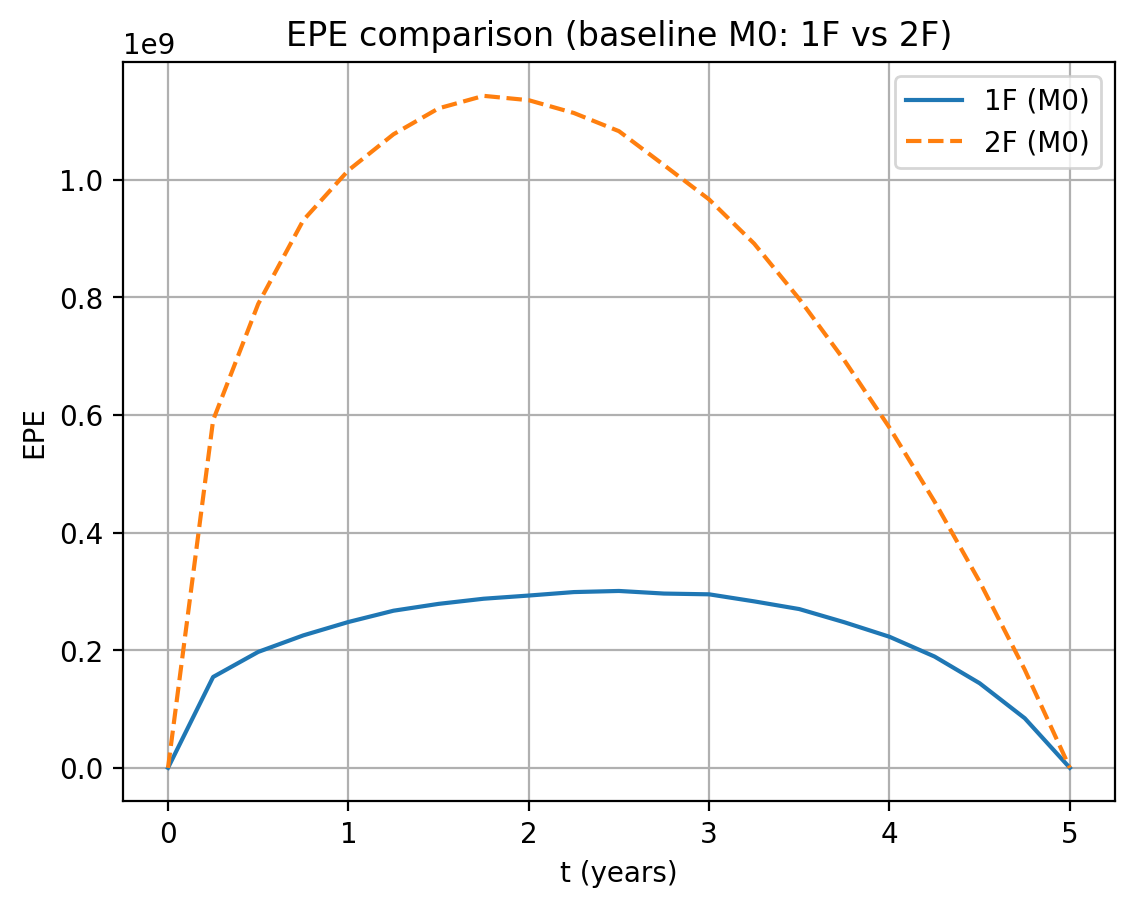}
\caption{Baseline-market EPE for the stylized WTI swap: one-factor Samuelson-type model (1F) versus a two-factor sensitivity specification (2F).}
\label{fig:wti_epe_1f2f}
\end{figure}

Commodity forward curves often require more than one factor to capture shape dynamics, so we repeat the market/credit decomposition under a simple two-factor curve specification in the spirit of \citet{SchwartzSmith2000}, adding a long-lived factor to the Samuelson-type short factor. Figure~\ref{fig:wti_epe_1f2f} compares baseline-market EPE under the one-factor versus two-factor exposure models, and Table~\ref{tab:wti_decomp_2f} reports the corresponding decomposition.
For CCVA attribution, the interaction share remains close to the one-factor result and indicates that credit and market channels cannot be cleanly separated additively. Applying the KL-robust WWR bound from Section~\ref{sec:method} to the same WTI swap loss samples yields a scenario-relative robust $\CCVAupper$ that increases monotonically with the dependence budget $\varepsilon$ (Table~\ref{tab:wti_wwr_eps}).

\begin{table}[!htbp]
\centering
\caption{WTI swap under a two-factor model (USD mm).}
\label{tab:wti_decomp_2f}
\begin{tabular}{lcc}
\toprule
Panel A: CVA (USD mm) & $M0$ & $M1$ \\
\midrule
Credit $\lambda_0$ (Current Policies) & 53.01 & 20.95 \\
Credit $\lambda_1$ (Net Zero 2050)    & 61.84 & 24.90 \\
\bottomrule
\end{tabular}

\vspace{0.8em}

\begin{tabular}{lrrrrr}
\toprule
Panel B: decomposition (USD mm) & $\Delta\mathrm{CVA}_{\text{credit}}$ & $\Delta\mathrm{CVA}_{\text{market}}$ & $\Delta\mathrm{CVA}_{\text{interaction}}$ & $\Delta\mathrm{CVA}$ & share \\
\midrule
2F (robustness) & 8.83 & -32.06 & -4.88 & -28.11 & 17.4\% \\
\bottomrule
\end{tabular}

\end{table}

\begin{table}[!htbp]
\centering
\caption{KL-robust WWR. Values are in USD millions and report $\CCVAupper$ as a function of $\varepsilon$.}
\label{tab:wti_wwr_eps}
\begin{tabular}{rrrr}
\toprule
$\varepsilon$ & $\mathrm{CCVA}^{\mathrm{ind}}$ & $\CCVAupper$  & WWR  \\
\midrule
0.01 & 2.94 & 3.97 & 1.03 \\
0.05 & 2.94 & 5.28 & 2.33 \\
0.10 & 2.94 & 6.28 & 3.33 \\
0.20 & 2.94 & 7.69 & 4.74 \\
\bottomrule
\end{tabular}

\end{table}
\FloatBarrier

\section{Hull--White 1F exposure model and swap valuation}\label{app:hw1f_swap}
Under the risk-neutral measure, the short rate $r_t$ follows the one-factor Hull--White model
\begin{equation}
dr_t = \bigl(\theta(t) - a r_t\bigr)\,dt + \sigma\, dW_t,
\label{eq:hw_sde}
\end{equation}
with mean-reversion $a>0$ and volatility $\sigma>0$. The drift function $\theta(t)$ is chosen so that the model exactly fits the input discount curve $P^M(0,T)$ at $t=0$.
Writing the market instantaneous forward rate as $f^M(0,t)=-\partial_t \log P^M(0,t)$, a standard curve-fitting choice is
\begin{equation}
\theta(t) = \partial_t f^M(0,t) + a f^M(0,t) + \frac{\sigma^2}{2a^2}\bigl(1-e^{-at}\bigr)^2,
\label{eq:hw_theta}
\end{equation}
which implies $P(0,T)=P^M(0,T)$ for all $T$ on the fitted curve.

In Hull--White 1F, zero-coupon bond prices are exponential-affine in the short rate:
\begin{equation}
P(t,T) = A(t,T)\,\exp\!\bigl(-B(t,T)\,r_t\bigr),
\qquad
B(t,T)=\frac{1-e^{-a(T-t)}}{a},
\label{eq:hw_bond}
\end{equation}
where $A(t,T)$ is deterministic given the fitted initial curve and $(a,\sigma)$. We use \eqref{eq:hw_bond} to compute discount factors at future times along simulated paths.

We consider a plain-vanilla fixed-for-floating swap with payment dates $T_1<\cdots<T_m$, accrual factors $\alpha_k$, fixed rate $K$, and notional $N$. In a single-curve setting, the time-$t$ present value of the fixed leg is
\[
PV^{\mathrm{fixed}}_t = N K \sum_{k=1}^m \alpha_k\, P(t,T_k),
\]
while the floating leg can be written as the difference of discount factors,
\[
PV^{\mathrm{float}}_t = N\bigl(1-P(t,T_m)\bigr),
\]
which is exact for a spot-starting swap with the next reset at $t$ under the benchmark's stylized convention. The swap mark-to-market is $V_t=PV^{\mathrm{float}}_t-PV^{\mathrm{fixed}}_t$ for a payer-fixed position.

Positive exposure on the simulation grid is defined as $E(t_i)=\max(V_{t_i},0)$, and expected positive exposure is $\mathrm{EPE}(t_i)=\mathbb{E}[E(t_i)]$ across Monte Carlo paths. These exposures feed into the discrete CVA approximation in Eq.~\eqref{eq:cva_discrete} and the loss samples used in the KL-robust WWR.

\end{document}